\documentclass[12pt]{article}
\usepackage{graphicx}
\usepackage{psfig}

\usepackage{scicite}

\usepackage{times}

\newenvironment{sciabstract}{%
\begin{quote} \bf}
{\end{quote}}

\newcounter{lastnote}

\title{Evolution of the Magnetic Field Structure of the Crab Pulsar}

\author
{Andrew Lyne,$^{1\ast}$ Francis Graham-Smith,$^{1}$ Patrick Weltevrede,$^{1}$\\
Christine Jordan,$^{1}$ Ben Stappers,$^{1}$ Cees Bassa,$^{1}$ Michael Kramer$^{1,2}$\\
\\
\normalsize{$^{1}$ Jodrell Bank Centre for Astrophysics, School of
Physics and Astronomy,}\\
\normalsize{University of Manchester, Manchester, M13~9PL, UK}\\
\normalsize{$^{2}$ MPI f\"ur Radioastronomie, Auf dem H\"ugel 69,
53121 Bonn, Germany}\\
\normalsize{$^\ast$To whom correspondence should be addressed; E-mail:  andrew.lyne@manchester.ac.uk.}
}

\date{}

\begin{document} 


\baselineskip24pt


\maketitle


\begin{sciabstract}
Pulsars are highly-magnetised rotating neutron stars and are
well-known for the stability of their signature pulse shapes, allowing
high-precision studies of their rotation.  However, during the past 22
years, the radio pulse profile of the Crab pulsar has shown a steady
increase in the separation of the main pulse and interpulse components
at 0.62$^{\rm o}\pm$0.03$^{\rm o}$ per century.  There are also
secular changes in the relative strengths of several components of the
profile.  The changing component separation indicates that the axis of
the dipolar magnetic field, embedded in the neutron star, is moving
towards the stellar equator.  This evolution of the magnetic field
could explain why the pulsar does not spin down as expected from
simple braking by a rotating dipolar magnetic field.
\end{sciabstract}


The Crab pulsar\cite{sr68} originated in the collapse of the core of a
massive star in AD~1054, an event that was visible on Earth through
the subsequent supernova explosion.  Presently rotating at 30 times a
second\cite{ccl+69}, intense beams of electromagnetic radiation from
the magnetic poles of the neutron star sweep across the
Earth\cite{gol68}, resulting in pulsed emission which has been
observed since 1969 at radio wavelengths and subsequently at optical,
X-ray and gamma-ray wavelengths. The pulse profile is related to the
shape of the beams which are determined by the magnetic field structure
of the underlying star as it rotates\cite{gol68}.  We have therefore
sought changes in the radio pulse profile in order to investigate any
evolution in the structure of the stellar magnetic field.

\begin{figure*}
\begin{center}
\includegraphics[width=12cm]{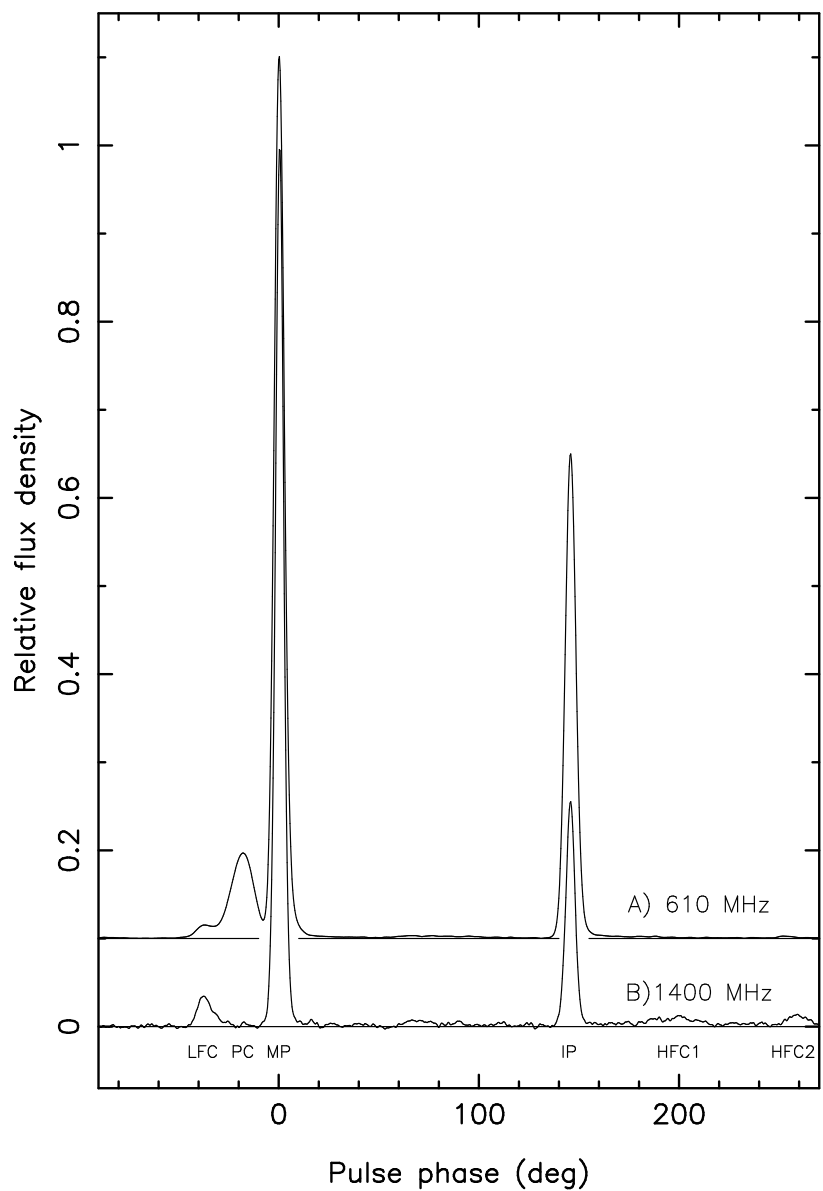}
\end{center}
\caption[]{The average pulse profiles of the Crab
pulsar.  A)~610~MHz and B)~1400~MHz. The locations of the features
discussed in the text are indicated below the profiles: the main pulse
(MP) and interpulse (IP) lie at rotational phases 0$^{\rm o}$ and
145$^{\rm o}$, while the precursor (PC) and Low-Frequency Component
(LFC) lie at $-$18$^{\rm o}$ and $-$37$^{\rm o}$, respectively.}
\label{fg:profiles}
\end{figure*}

High-quality daily observations of the Crab pulsar (PSR B0531+21,
J0534+2200) have been made since 1991 with the 13-m radio telescope at
Jodrell Bank Observatory at a frequency of 610 MHz.  These
measurements are supplemented by less-frequent observations using the
76-m Lovell Telescope at the higher frequency of 1400 MHz, designed to
monitor dispersion measure.  These data comprise part of a consistent
set of pulse timing observations at Jodrell Bank\cite{lps88,lps93}
that have provided a complete record of the rotation of the pulsar
since 1984, and are available as an ephemeris to other
observers\cite{ljr13}. The radio pulse profile
(Fig.~\ref{fg:profiles}) consists of a pair of components, the main
pulse (MP, at 0$^{\rm o}$) and the interpulse (IP, at 145$^{\rm o}$),
which are closely associated with the components of the profiles
observed in the high-energy regimes from optical to TeV
gamma-rays\cite{aaa+10b}.  At the comparatively low radio frequency of
610 MHz there is also a steep-spectrum third component known as the
precursor (PC, at $-$18$^{\rm o}$).  This component is not detectable
at 1400 MHz, although another, previously-identified component, the
``low-frequency component'' (LFC, at $-$37$^{\rm o}$)\cite{mh96}, is
seen at both frequencies. The components HFC1 (at 200$^{\rm o}$) and
HFC2 (at 260$^{\rm o}$) reported before\cite{mh96} also appear in the
1400-MHz profile, but are too weak to be addressed in this study.

\begin{figure*}
\begin{center}
\includegraphics[width=12cm]{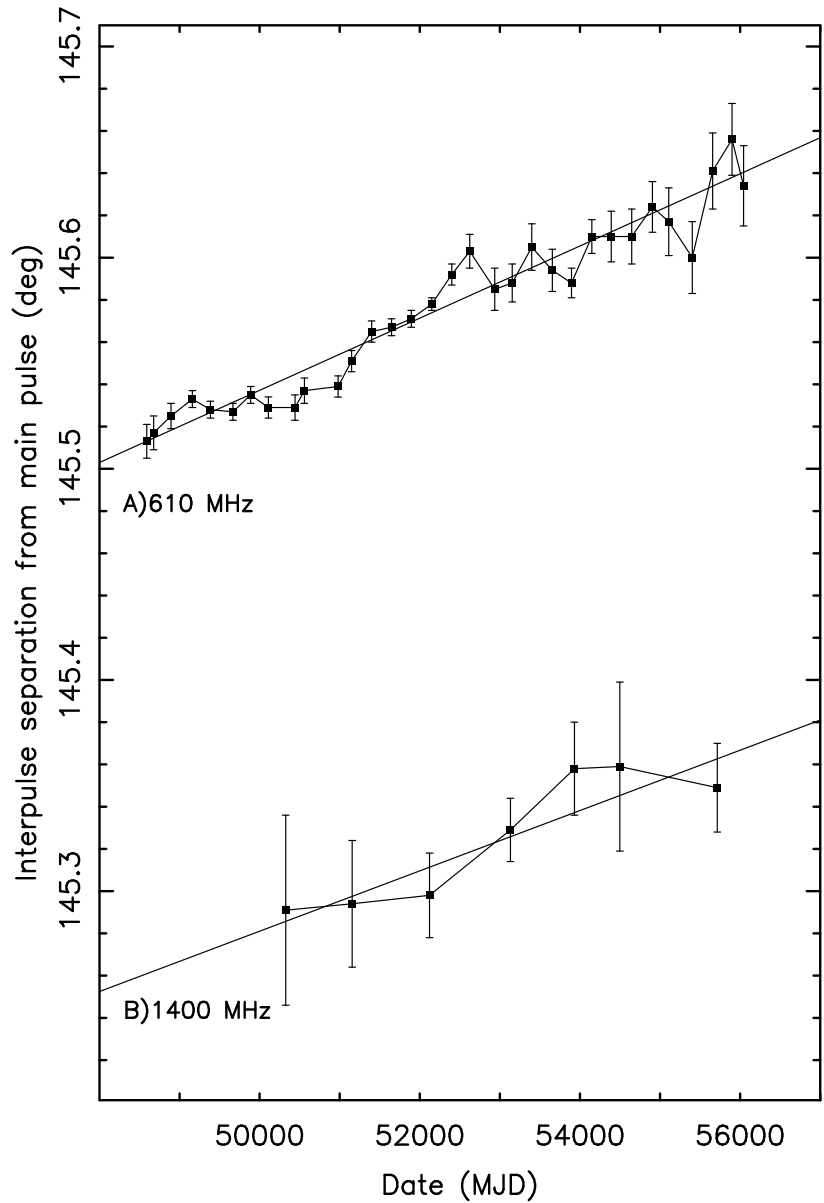}
\end{center}
\caption[]{The rotational separation of the IP
from the MP.  A) 610~MHz and B) 1400~MHz.  Data at 610~MHz are
displayed as mean values averaged over 500 days, calculated at 250-day
intervals, while data at 1400~MHz are mean values averaged over 2000
days, calculated at 1000-day intervals. }
\label{fg:spacing1}
\end{figure*}

\begin{figure*}
\begin{center}
\includegraphics[width=12cm]{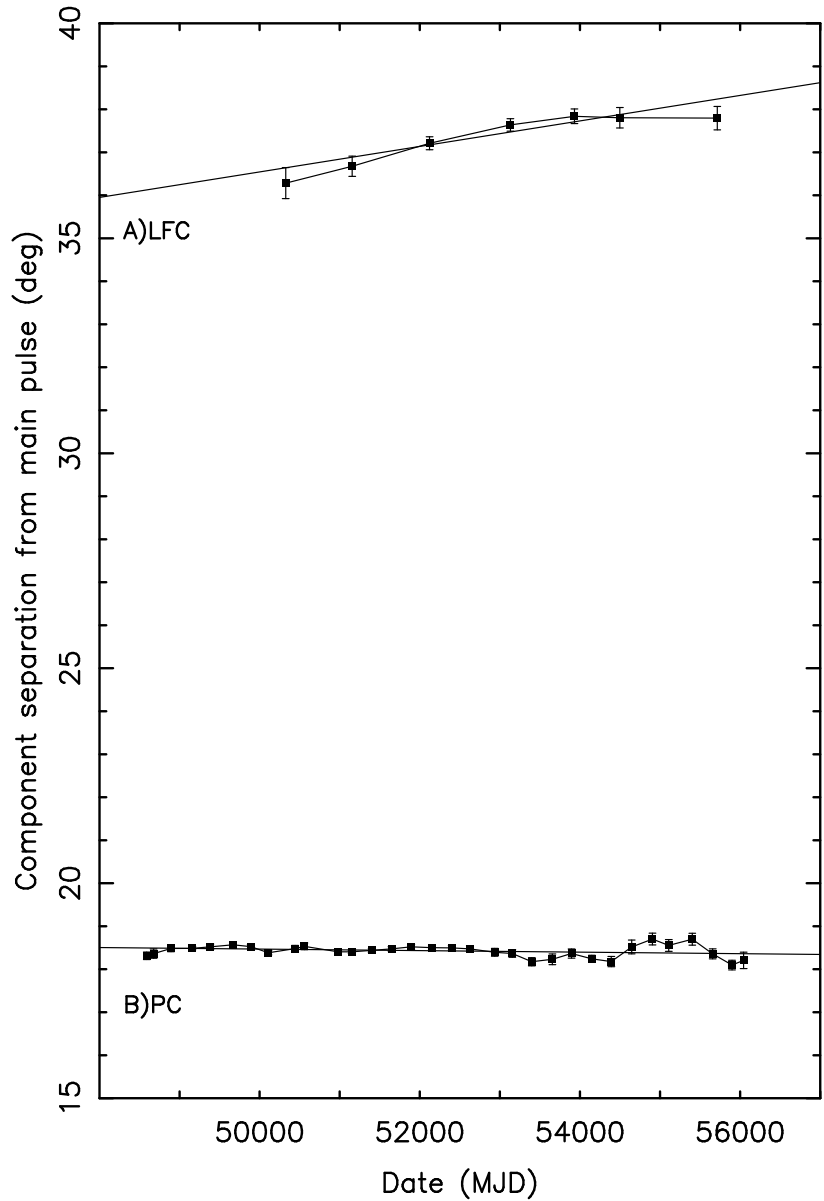}
\end{center}
\caption[]{The rotational separation of
components of the Crab pulse profile from the MP. A) the LFC at
1400~MHz and B) the PC at 610~MHz. Data at 610~MHz are displayed as
mean values averaged over 500 days, calculated at 250-day intervals,
while data at 1400~MHz are mean values averaged over 2000 days,
calculated at 1000-day intervals. }
\label{fg:spacing2}
\end{figure*}

\begin{figure*}
\begin{center}
\includegraphics[width=12cm]{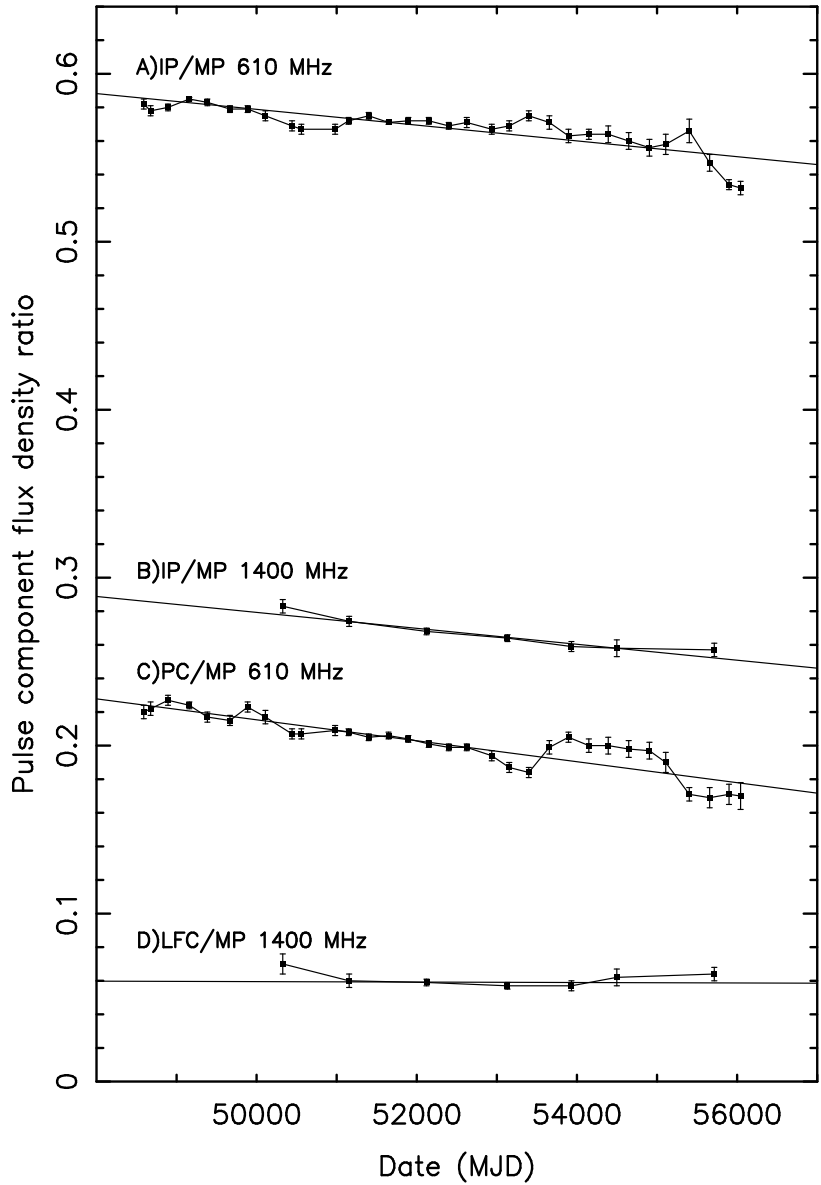}
\end{center}
\caption[]{Relative flux densities of Crab pulse profile
components.  Flux densities are given relative to that of the MP at
the same frequency: A) IP at 610~MHz B) IP at
1400~MHz and C) PC at 610~MHz and D) LFC at 1400~MHz. Data at 610~MHz
are displayed as mean values averaged over 500 days, calculated at
250-day intervals, while data at 1400~MHz are mean values averaged
over 2000 days, calculated at 1000-day intervals. }
\label{fg:flux}
\end{figure*}

The temporal evolution of the separation between the MP and other
components\cite{meth13} can be seen in Fig.~\ref{fg:spacing1}
(IP) and Fig.~\ref{fg:spacing2} (LFC and PC), while variations in the
relative integrated flux densities of the pulse components are shown
in Fig.~\ref{fg:flux}.  Secular changes are observed in all three
diagrams.
The component separations and relative flux densities, together with
their rates of change measured over the 22 years, are presented in
Table~\ref{ta:psr_params}.  In summary, the most notable features are
that:

(i) the separation of the IP from the MP at 610~MHz
has increased by 0.14$^{\rm o}$, corresponding to $13\pm1\,\mu$s
(Fig.~\ref{fg:spacing1}), together with a consistent change at 1400 MHz.

(ii) the separation of the LFC from the MP
at 1400~MHz has increased by 2.4$^{\rm o}$, corresponding to
$220\pm40\,\mu$s (Fig.~\ref{fg:spacing2}A).

(iii) the flux density of the IP has decreased relative to
that of the MP by $\sim 6$\% at 610~MHz and by $\sim 13$\% at
1400~MHz (Fig.~\ref{fg:flux}A and Fig.~\ref{fg:flux}B).

(iv) the flux density of the PC at 610~MHz has
decreased relative to that of the MP by $\sim 13$\%
(Fig.~\ref{fg:flux}C).

\begin{table*}[bt]
\begin{center}
\caption{The rotational positions and integrated flux densities
of the interpulse (IP), precursor (PC) and low-frequency component
(LFC) of the radio profile of the Crab pulsar.  Values are given in
columns 2 and 4 relative to the MP at epoch MJD=53,000 at 610~MHz and
1400~MHz. Columns 3 and 5 contain the rates of change of these two
quantities respectively.  Standard ($1\,\sigma$) errors are given in
parentheses after the values and are in units of the least significant
quoted digit.}
\begin{tabular}{lccll}\\
\hline
\hline
         & Position & Rate of Change  & Flux Ratio & Rate of Change\\ 
         &  ($^{\rm o}$)     &  ($^{\rm o}$/century)  &   &  (/century) \\
\hline
 & & & & \\
IP (610 MHz) & 145.588(2)  & +0.62(3) & 0.5648(6) & $-$0.172(8) \\
IP (1400 MHz) & 145.324(9) & +0.5(2) & 0.265(1) & $-$0.17(3) \\
PC (610 MHz) & $-$18.415(17) & $-$0.6(3) & 0.1967(6)  & $-$0.226(11) \\
LFC (1400 MHz) & $-$37.43(7) & +11(2) & 0.059(1) & $-$0.005(33) \\
 & & & & \\
\hline
\end{tabular}
\label{ta:psr_params}
\end{center}
\end{table*}

The origins of the pulse components are within a co-rotating
magnetosphere which is embedded in the neutron star and extends almost
to the light cylinder, the cylindrical surface at which the
co-rotation velocity would reach the velocity of light.  The source of
the PC is believed to be located over a magnetic pole, at a small
fraction of the distance to the light cylinder. The MP and IP
components of the high-energy (optical, X-ray and gamma-ray) profile
originate in two high-altitude gaps in the magnetosphere, at the
boundary between magnetic field lines which close within the light
cylinder and the open field lines which are closer to the magnetic
poles. This emission occurs in an extended region along each gap,
while the combined effect of propagation time along the gap and
relativistic effects concentrates the observed source in to a
caustic\cite{mor83}. While this broadly describes the emission, the
exact distribution depends on the specific emission model
(e.g.~\cite{chr86a,ry95,mh04a,dr03,rw10}). In the following, we
construct a geometrical interpretation for the secular increase in the
separation of the MP and IP radio emission. This is possible because
the radio components are closely associated with the high energy
emission, thereby suggesting that high energy models can be used to
interpret the observed evolution of this separation.

The observed high-energy profile is a cut across a hollow conical
surface of emission over one or both magnetic poles.  The shape of the
cone is determined by the inclination $\alpha$ of the magnetic axis to
the rotation axis, and the position of the cut depends on the viewing
angle $\zeta$, which is measured from the rotation axis.  There is
good evidence from the geometry of the X-ray torus in the pulsar wind
nebula seen around the
pulsar \cite{nr08} that the viewing angle $\zeta$ is close to 63$^{\rm
o}$.  The observed secular increase of the separation between the IP
and MP might be due to a change in viewing geometry due to
precession\cite{ja01} (on a
timescale much larger than the 2.3\% of the 960-year lifetime of the
pulsar that our observations cover), or in $\alpha$ (i.e alignment or
counter alignment of the magnetic and rotation axes) or in the
location within the magnetosphere of the source of emission.  While a
long-period free precession of the neutron star might provide an
appropriate change in the beam geometry, not only are there
theoretical arguments why it should not occur\cite{sha77,swc99}, but
no pulsar has been shown to display the phenomenon
(e.g.~\cite{lhk+10}). Here we demonstrate that a change in
$\alpha$ will result in a self-consistent interpretation.

The shape of the beam of the Crab pulsar has been modelled by several
authors (e.g.~\cite{dr03,hsdf08,wrwj09,dqw12}), resulting in a range
of estimates of $\alpha$ between 45$^{\rm o}$ and 70$^{\rm o}$.
Nevertheless, a common feature of these models is a prediction that an
increasing $\alpha$ will result in an increasing peak separation
(e.g. Figs.~5 and 2 in \cite{wrwj09}, from which it can be deduced
that the peak separation is expected to change at a rate of about one
degree per degree change in $\alpha$), although the exact relationship
is somewhat model-dependent.  The observed secular increase of the
separation therefore indicates that the dipole axis is moving towards
orthogonality.  The observed component separation is increasing from
145.5$^{\rm o}$ towards the symmetrical value of 180$^{\rm o}$ at a
rate of 0.62$^{\rm o}$ per century (Fig.~\ref{fg:spacing1}). We
therefore interpret our observation as an increase of $\alpha$ at a
comparable rate, at which rate the total change would have been about
6$^{\rm o}$ during the lifetime of the pulsar so far.

It is difficult to relate the rapid apparent motion of the LFC, whose
origin is not understood, to this geometric evolution in any detail.
The changes in relative flux densities of the components are easier to
understand because all the sources are highly coherent and are probably
narrow beamed\cite{lg12a}, so that small structural magnetospheric
changes might cause large effects on the component flux densities.

The dipole magnetic field is embedded in the superconducting interior
of the neutron star, either in the inner crust or in the core, and
only a slow evolution can be expected.  An evolution towards alignment
rather than orthogonality is expected for a simple magnetic dipole.
However, it has been pointed out that an evolution towards
orthogonality may be expected because of the torque developed by the
return current in the neutron star surface (e.g.~\cite{bn07}). We also
note that the increasing slowdown torque due to increasing $\alpha$
could at least in part explain the observed braking index value of
2.50\cite{lps93}, rather than 3.0 expected for classic magnetic dipole
braking\cite{macy74}, or may be related to the occurrence of glitches
seen in the pulsar\cite{le97,ah97}. If the sole departure from
such classic slowdown, in which the rate of change of rotation rate
$\dot{\nu} \propto \nu^3 {\rm sin}^2 (\alpha)$, is a secular change in
$\alpha$ at a rate $\dot{\alpha}$, the observed braking index $n$ is
given by $ n = 3 + 2 \nu / \dot{\nu} \times \dot{\alpha} / {\rm
tan}(\alpha)$.  Within the limitations of our model, the observed
braking index can be explained if $\alpha$ increases at 0.6$^{\rm o}$
per century, which is remarkably close to the rate of change in
$\alpha$ required to explain the secular change in the pulse
separation.

These observations provide evidence for a progressive
change in the magnetic inclination of an isolated pulsar. The precise
measurement of the small change in pulse profile, which leads to this
conclusion, has depended on a long and consistent series of
observations of one of the youngest pulsars.  It is unlikely that
comparable measurements can be made on any other pulsar in the near
future.




Pulsar research at JBCA is supported by a Consolidated Grant from the UK
Science and Technology Facilities Council.

\noindent {\bf Supplementary Materials}\\
www.sciencemag.org\\
Materials and Methods\\
References ({\it20,29,30})

\clearpage
\setcounter{page}{1}

\noindent{\bf Materials and methods}

The method of determination of the pulse shape parameters from the
observed pulse profiles used here was similar to that described
earlier ({\it20}).  Each of the two pulse profiles shown in
Fig.~1 was inspected and fitted with up to five Gaussian
components, as required to provide satisfactory descriptions of the
profiles. Two Gaussian components were used to describe each of the MP
and IP and a single one for either the PC (at 610~MHz) or LFC (at
1400~MHz).  Components which initially had these amplitudes, widths
and relative positions were then fitted to each observed profile, by
adjusting the values of the parameters in a least-squares
minimisation process to produce a synthetic profile. From this
synthetic profile, values of parameters such as component integrated
flux densities and separations were determined and presented in
Table~1 and Figs.~2-4.
Note that this procedure has the virtue of applying a quasi-optimum
filter to the data in order to minimise the effects of high-frequency
noise in the profiles on the values of the parameters.

The radio pulse profile of the Crab pulsar at around 610~MHz suffers
from a variable amount of multipath scattering from plasma in the Crab
Nebula ({\it29,30}), causing significant asymmetric distortion
of the profile. To first order this has no effect upon the measured
component separations, since all components are delayed by the same
amount.  However, because of subtle differences in the underlying
component structures, trials showed that increasing the amount of
smoothing due to scattering resulted in small systematic changes in
the measured separation of the MP and IP, amounting to
(W$_{50}-$6.0)$\times$0.035$^{\rm o}$, where W$_{50}$ is the FWHM of
the MP in degrees.  This correction was applied to the measured values
of MP and IP separation.  In about 5\% of the observations, strong
scattering caused W$_{50}$ to increase to greater than 7$^{\rm o}$, in
which cases the component positions were more poorly defined and the
observations were discarded.  For the remaining data, the average
magnitude of the applied correction was 0.008$^{\rm o}$, notably
smaller than the secular variations reported in this paper.


\clearpage
\setcounter{page}{10}

\setcounter{figure}{0}

\end{document}